\RequirePackage{fix-cm}
\documentclass[smallcondensed]{svjour3}     
\smartqed  
\usepackage{graphicx}
\usepackage{amsmath}
\usepackage{graphicx}
\usepackage{amssymb}
\usepackage{color}
\usepackage{bm}
\usepackage{authblk}
\journalname{Transport in Porous Media}
\begin{document}

\title{Transport of polymer particles in a oil-water flow in porous media: enhancing oil recovery\thanks{This work was partially supported by Statoil, through Akademiaavtalen, the Norwegian Academy of Science and Letters and Statoil through the VISTA AdaSim project No. 6367.}
}

\titlerunning{Transport of polymer particles in oil-water flows in porous media}        

\author{M.A. Endo Kokubun \and
        F.A. Radu \and
        E. Keilegavlen \and
        K. Kumar \and
        K. Spildo
}


\institute{M.A. Endo Kokubun \at
              Department of Chemistry, University of Bergen, All\'{e}gaten 41, 5007 Bergen, Hordaland, Norway \\
              Tel.: +47 55 58 34 44\\
              \email{max.kokubun@uib.no}           
           \and
           F.A. Radu \at
              Department of Mathematics, University of Bergen, All\'{e}gaten 41, 5007 Bergen, Hordaland, Norway\\
              \email{florin.radu@uib.no}
           \and
           E. Keilegavlen \at
              Department of Mathematics, University of Bergen, All\'{e}gaten 41, 5007 Bergen, Hordaland, Norway\\
              \email{eirik.keilegavlen@uib.no}
           \and
           K. Kumar \at
              Department of Mathematics, University of Bergen, All\'{e}gaten 41, 5007 Bergen, Hordaland, Norway   \\        
              \email{kundan.kumar@uib.no}
           \and
           K. Spildo \at
              Department of Chemistry, University of Bergen, All\'{e}gaten 41, 5007 Bergen, Hordaland, Norway        \\   
              \email{kristine.spildo@uib.no}
}

\date{Received: date / Accepted: date}

\maketitle

\begin{abstract}
We study a heuristic, core-scale model for the transport of polymer particles in a two phase (oil and water) porous medium.
We are motivated by recent experimental observations which report increased oil 
recovery when polymers are injected after the initial waterflood.
The recovery mechanism is believed to be microscopic diversion of the flow, 
where injected particles can accumulate in narrow pore throats and clog it, in 
a process known as a log-jamming effect.
The blockage of the narrow pore channels lead to a microscopic diversion of the 
water flow, causing a redistribution of the local pressure, which again can 
lead to the mobilization of trapped oil, enhancing its recovery.
Our objective herein is to develop a core-scale model that is consistent with
the observed production profiles.
We show that previously obtained experimental results can be qualitatively explained by a simple two-phase flow model with an additional transport equation for the polymer particles.
A key aspect of the formulation is that the microscopic heterogeneity of the rock and a dynamic altering of the permeability must be taken into account in the rate equations.

\end{abstract}

\section{Introduction}
\label{intro}

After initial waterflood, a large quantity of oil is still left inside the 
reservoir, with remaining oil ratios frequently exceeding $50\%$ of the total 
amount \cite{lake1992}.
Techniques to recover this remaining oil are known as tertiary recovery methods (enhanced oil recovery) and many different approaches have been proposed \cite{shengBook} (in-situ combustion, steam injection, polymer injection, microbial enhanced recovery, etc).
The choice of one or another recovery method depends on a myriad of factors, ranging from economical costs, geological characteristics of the reservoir, thermophysical properties of the remaining oil, geographical location of the reservoir, etc.

\begin{table}[h!]
	\centering
	\caption{Nomenclature}
	\label{symbols}
	\begin{tabular}{|l |l|}
	\hline\noalign{\smallskip}
	{\bf Nomenclature}  & \\
	                    & \\
	$A_1, A_2$          & constants for residual oil saturation change \\
	$c_l$               & mass concentration of polymer particles (kg$/$m$^3$)\\
	${\bf D_l}$         & Diffusion matrix of polymer particles in water (m$^2/$s) \\
	$d_p$               & polymer particle diameter (m) \\
	$\vec{g}$           & gravity vector (m$/$s$^2$)\\	
	${\bf K}$           & absolute permeability tensor (m$^2$) \\	
	$K_B$               & Boltzmann constant (m$^2$kg$/$s$^2$K) \\
	$k_l$               & constant rate of clogging (m$^{-1}$) \\
	$k_r$               & constant rate of unclogging (s$^{-1}$) \\	
	$k_{r\alpha}$       & relative permeability of phase $\alpha$ \\
	$n_o,n_w$           & exponents for the relative permeabilities\\
	$p_\alpha$          & pressure of phase $\alpha$ (Pa) \\
	$R$                 & reaction rate (kg$/$m$^3$s)\\
	$s_\alpha$          & saturation of phase $\alpha$ \\
	$T$                 & temperature (K) \\
	$\vec{u}_\alpha$    & velocity of phase $\alpha$ \\	
						& \\
    {\it Greek letters} & \\						
	$\beta$             & proportionality factor for reaction rate\\
	$\gamma$            & constant for permeability change \\	
	$\delta$            & heterogeneity factor \\
	$\lambda_\alpha$    & mobility of phase $\alpha$ (m~s$/$kg)\\
	$\mu_\alpha$        & viscosity of phase $\alpha$ (kg$/$m~s)\\
	$\rho_\alpha$       & mass density of phase $\alpha$ (kg$/$m$^3$) \\	
	$\sigma$            & volumetric concentration of accumulated particles\\
	$\tau$              & tortuosity \\
	$\phi$              & porosity \\
	$\varphi$           & clogging rate \\
	$\psi$              & unclogging rate \\
\noalign{\smallskip}\hline
	\end{tabular}
\end{table}

Injection of low salinity water into porous cores has been shown to decrease the rock permeability due to the lifting, migration and subsequent plugging of pores by fine particles \cite{li2011,morrow2011,sheng2014}.
If the plugging is severe, the pore can be completely clogged, such that flow is no longer allowed.
It is possible that the particles released from the rock due to the injection of low salinity water have sizes comparable to the pore diameter.
If that is the case, straining will be the main mechanism of plugging, i.e., a single particle can clog a pore.
This leads to a permeability decline in swept zones, which may increase recovery of oil due to a improved mobility control of water.
Although this method has the advantage that the particles are generated in-situ, controlling the amount of released fine particles is difficult.
Change in wettability also plays a significant role in low-salinity water injection \cite{abbas2016,bedrikovetsky2017}.

Injecting nanoparticles along with water into the reservoir has the advantage of controlling the quantity of particles in the core.
There are several different types of nanoparticles that can be used in such process of enhancement of oil recovery, but most of the studies focus on nanoparticles that can change the wettability of the rock, thus changing capillary pressure curves and improving oil recovery \cite{elamin2013}.
In the case of nanoparticles that interact with the rock (through electrostatic potential, for example \cite{bennacer2017,ray2012}), attachment rates must be considered in order to account for the deposition of the particles at pore walls.

\begin{figure*}[h!]
\begin{center}
\includegraphics[width=\linewidth]{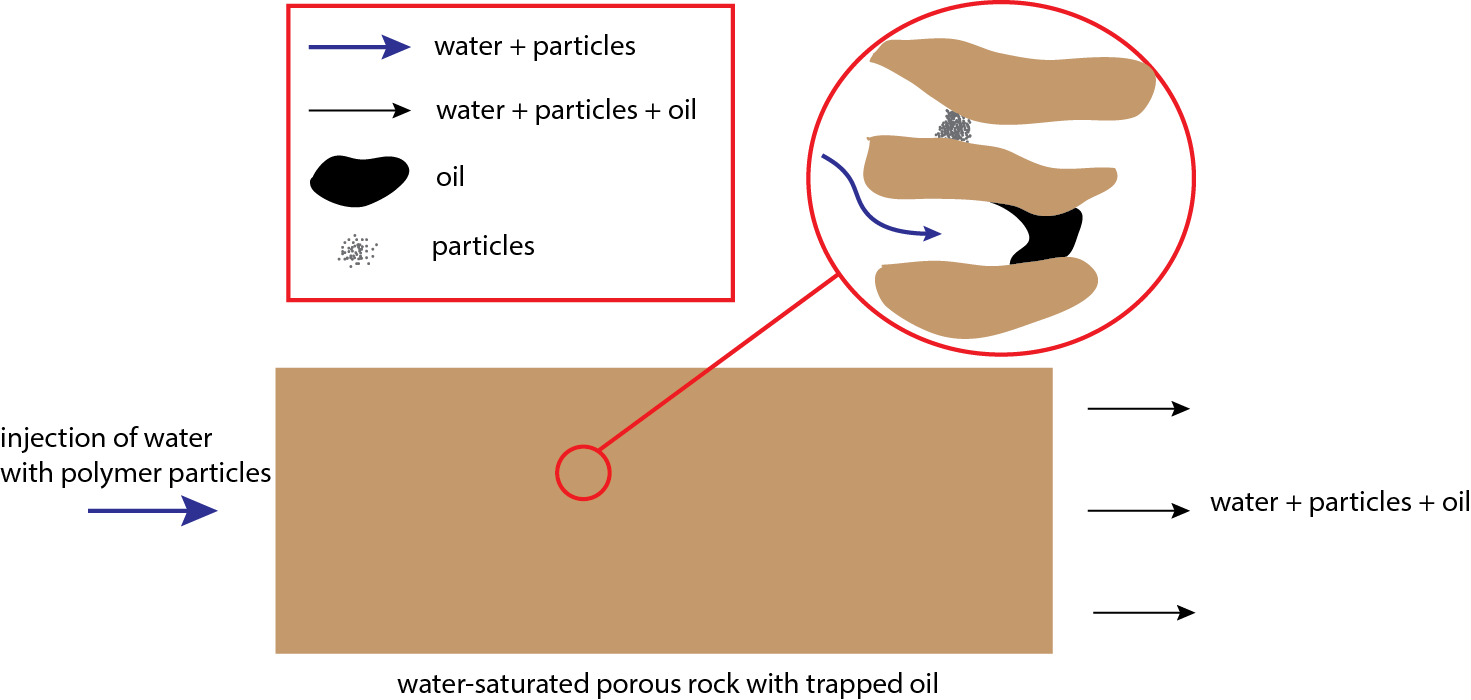}
\end{center}
\caption{Polymer particles injected along with water flow into a porous rock with trapped oil. The clogging of a pore channel and the consequent water flow diversion leading to the recovery of initially trapped oil is shown in the detail.}
\label{fig1}
\end{figure*}

The injection of polymer particles along with water aims, on the other hand, at recovering the trapped oil through a mechanism of microscopic diversion of the flow.
The injected polymer particles can accumulate in narrow pore throats, leading to a log-jamming effect that essentially diverts the water flow.
This mechanism was proposed in order to explain recently obtained experimental results \cite{spildo2009,spildo2010}.
A schematic is shown in Fig. \ref{fig1}.
When a pore throat is clogged, the water is diverted towards the oil-containing neighbouring pore channel.
If the pressure increase in the oil-containing channel is large enough, the trapped oil is mobilized and can be recovered.
After oil is mobilized, conditions may favour unclogging of the pore throat.
Since the characteristics of the polymer particles are such that they don't 
change the wettability of the rock, microscopic diversion of the flow is one of the main responsible for enhancing oil recovery during polymer particle injection.

Polymer particles are complex particles which are formed by mixing a low 
concentration aqueous-polymer solution with a crosslinker. 
Effects that can contribute to oil mobilization 
include viscosity changes of the polymer-carrying water, absorption, incomplete 
complexation of polymers and crosslinkers, which can lead to polymer- and 
particle-carrying with different flow properties, and log-jamming.
If the complexation between crosslinkers and polymer is incomplete, some pure polymer 
might be injected along with the polymer particles.
If that is the case, the governing equations should distinguish between two  
polymer-carrying and particles-carrying water, as these have different effects 
on flow properties (in addition to a conservation equation for the oil phase).
Nevertheless, pore network models have shown that log-jamming plays a more 
prominent role in permeability reduction and variation in the core pressure 
difference, when compared to viscosity change, polymer straining and adsorption 
\cite{bolandtaba2009,bolandtaba2011}.
Therefore, as a first approach one can construct a simple model which neglects incomplete complexation, straining and adsorption in the governing equations, thus neglecting polymer-carrying water effects.
This first approach can then be used as a starting point for constructing more realistic models, which take into account the complex physical interactions of this problem.

Some recent experimental results \cite{spildo2009,spildo2010} have shown that injection of polymer particles along with water into a rock containing trapped oil exhibit some different characteristics when compared to other particle-carrying systems (such as fines migration or nanoparticle injection). 
For example, even though the enhancement in oil recovery is prominent, measurements of pressure drops along the core have shown that some of the tested cores exhibited a very small final permeability damage.
This indicates that unclogging occurred after oil mobilization took place.
Moreover, measurements in Berea sandstone cores showed negligible enhancement 
in oil recovery due to polymer particle injection, which suggests that some 
cores may be more suitable to polymer particle injection as an enhanced oil 
recovery method than others.
It is our intention, in this paper, to develop a simple mathematical model that correctly mimics the experimental results described in \cite{spildo2009,spildo2010}, by identifying the relevant physical mechanisms present in this problem.

Due to the local nature of the relevant processes that influence the log-jamming effect (pressure difference in each pore, different particle concentrations and velocities, etc), and the intrinsic nature of upscaled models (which consider averaged values for the quantities of interest in a representative volume that contains a statistically significant amount of pores), no rigorous modelling of particle accumulation in two-phase flow in porous media are available.
Therefore, we study a simple heuristic, core-scale model that takes into account the transport of polymer particles diluted in the water phase and the mechanisms of clogging and unclogging of pores.
We consider a non-equilibrium reaction for the accumulation and release of particles at the narrow pore throats.
A key aspect that must be taken into account is that the heterogeneity of the 
rock, specifically in the pore throat radius distribution, has a significant 
role on the particle accumulation process 
\cite{spildo2009}.
This can be explained by pore-scale acceleration and inertia effects for 
particle-carrying water, which will be more pronounced in heterogeneous rocks 
\cite{anna2017}.
Our model accounts for this effect by relating the reaction coefficient in the 
model for accumulation to the distribution of pore throat sizes.
We further include reductions in the residual oil saturation due to the 
presence of particles.

This papers is structured as follows.
In Section \ref{sec.model} we present the model, introducing the rate equation, with an emphasis on the conceptual model of log-jamming.
In Section \ref{sec.num} we present some numerical solutions to the model.
First we present a representative case in order to highlight the main characteristics of the model and then we compare our results with the available experimental data.
We finalize the paper with the Conclusions.
In the Appendix we present the heterogeneity parameter for the cores utilized in previous experiments \cite{spildo2009}.

\section{Model formulation}
\label{sec.model}

We consider that polymer particles are transported only by the water phase and do not interact with the oil phase, i.e., mass transfer between phases and possible interface effects are neglected.
The water and oil phase mass conservation equations are given by
\begin{equation}
\frac{\partial}{\partial t}\left(\phi s_\alpha\right)
+
\nabla\cdot\vec{u}_\alpha
= 0,
\label{eq.s}
\end{equation}
where $\alpha\in\{w,o\}$, denotes the water and oil phase, respectively, $\phi$ is the porosity, $s_\alpha$ the saturation of phase $\alpha$ and $\vec{u}_\alpha$ is the Darcy velocity of phase $\alpha$.
The flow is considered incompressible, as we do not consider the presence of a gas phase.
The velocity of phase $\alpha$ is given by Darcy's law
\begin{equation}
\vec{u}_\alpha
=
-{\bm K}\lambda_\alpha
\left(\nabla p_\alpha - \rho_\alpha\vec{g}\right),
\label{eq.ua}
\end{equation}
where $\lambda_\alpha = k_{r\alpha}/\mu_\alpha$ is the mobility of phase $\alpha$, with viscosity $\mu_\alpha$ and relative permeability $k_{r\alpha}$.
The relative permeability depends on the saturation of phase $\alpha$, while the absolute permeability tensor ${\bm K}$ depends on the rock properties.

The size of the injected particles vary in a narrow range around $100~nm$.
However, when experimentally preparing the polymer particles, the aqueous solution is filtered prior to injection \cite{spildo2010}.
Therefore, we assume that there is only a single size for the injected particles.
The polymer particles are transported by the water phase and can accumulate in narrow pore throats.
Denoting by $c_l$ the mass concentration of polymer particles, the advection-diffusion transport equation is given by
\begin{equation}
\frac{\partial}{\partial t}
\left(
\phi s_w c_l + \rho_l\sigma
\right)
+
\nabla
\cdot
\left(
\vec{u}_w c_l
-
{\bm D}_l \phi s_w\nabla c_l
\right)
=
0,
\label{eq.cl}
\end{equation}
where $\sigma$ is the volumetric concentration of polymer particles that have been trapped due to log-jamming and $\rho_l$ is the mass density of the polymer particles.
We consider Fick's law for diffusion of mass, with ${\bm D}_l$ as the diffusion matrix of polymer particles.

The volumetric concentration of accumulated particles $\sigma$ is obtained from the following rate equation
\begin{equation}
\rho_l\frac{\partial \sigma}{\partial t}
=
R,
\label{eq.sigmal}
\end{equation}
where $R$ is a non-equilibrium rate function that takes into account accumulation and release of particles in narrow pore throats.
The general nonlinear form of $R$ is given by
\begin{equation}
R = \beta(c_l,\sigma)(\varphi(c_l,\sigma) - \psi(c_l,\sigma)),
\label{eq.r}
\end{equation}
where $\beta(c_l,\sigma)$ is a proportionality factor and $\varphi$ and $\psi$ are the accumulation and removal functions of the particles.
The proper form of the reaction rate will be discussed in the next Section.

\subsection{Log-jamming and particle release}
\label{sec.log}

As far as the authors are aware, no rigorously derived expression for $R$ exists in the literature.
One of the major difficulties associated with its derivation is the fact that log-jamming is a process which intrinsically depends on pore-scale physics.
Therefore, a comprehensive model would have to be derived from the analysis of the conservation equations at the pore-scale, and then upscaled in order to obtain the reaction rate as a function coupled to the pore-scale physics \cite{carina2016,carina2017,muntean2017}.
To derive a rigorous upscaled model for two-phase flow, including the transport of particles, is outside the scope of this work.
Instead, we chose a heuristic construction for the reaction rate based on the a priori knowledge of the main physical aspects of the problem.

When the particles clog a pore, water flow is diverted, and the diverted water 
may mobilize initially-trapped oil located in neighbouring pores.
After clogging a pore, the accumulated particles may eventually be remobilized, 
unclogging the pore and restoring the water flow in it.
The clogging particles are brought to the pore by advection of water, such that we consider that the accumulation of particles is proportional to the convective flux of the particles $u_w c_l$.
We consider that when a pore channel is clogged, the flow is completely diverted to neighbouring channels, such that there is no accumulation of particles behind the clog.
This implies that there is no filtration of water through the accumulated particles.

\begin{figure*}[h!]
\begin{center}
\includegraphics[width=0.5\linewidth]{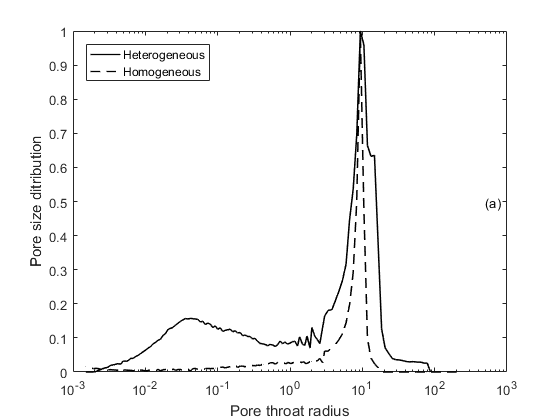}
\hspace{-0.5cm}
\includegraphics[width=0.5\linewidth]{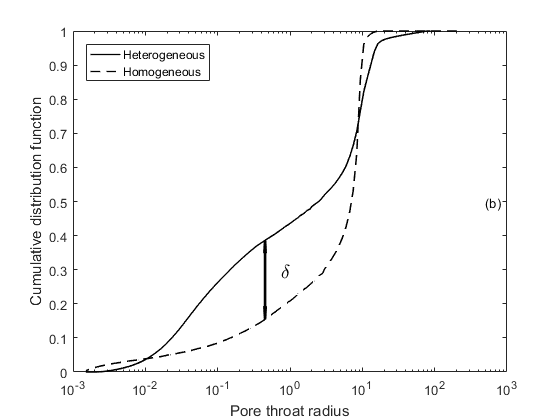}
\end{center}
\caption{(a) Pore-size distributions for a heterogeneous and a homogeneous core; and (b) their respective cumulative distribution functions.}
\label{fig.psd}
\end{figure*}

Experimental results indicate that cores with a higher degree of microscopic heterogeneity tend to have a higher accumulation of particles, which leads to higher recovery rates \cite{spildo2009}.
This can be explained by the fact that when water flows from a large to a 
narrow pore throat, it accelerates.
Due to inertia effects, the particles accumulate at the pore entrance and if enough accumulate, they clog the channel.
Therefore, we introduce a macroscopic measure of the microscopic rock heterogeneity into the rate of log-jamming.

Consider the two pore size distributions shown in Fig. \ref{fig.psd}(a).
The bimodal distribution, shown as a solid line, represents a heterogeneous core (taken from \cite{spildo2009}), with two predominant pore throat radii.
The unimodal distribution, shown as a dashed line, represents a homogeneous core (Berea).
Their cumulative distribution functions are shown in Fig. \ref{fig.psd}(b).
We consider the Kolmogorov-Smirnov statistics, which is a measure of the maximum separation between the two cumulative distribution curves, as an indicator of the degree of microscopic heterogeneity.
Denoting the maximum separation as $\delta \in [0,1]$, we can see that $\delta=0$ represents a homogeneous rock, whereas $\delta = 1$ represents a highly heterogeneous rock.
Since log-jamming occurs when particle-carrying water flows from the large pores to the narrow pores, $\delta$ measures the strength of log-jamming in the core.

\begin{figure*}[h!]
\begin{center}
\includegraphics[width=0.3\linewidth]{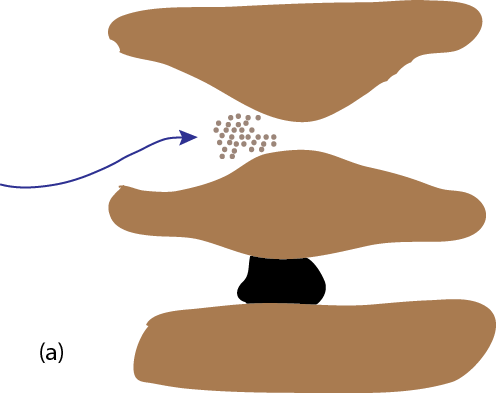}
\includegraphics[width=0.3\linewidth]{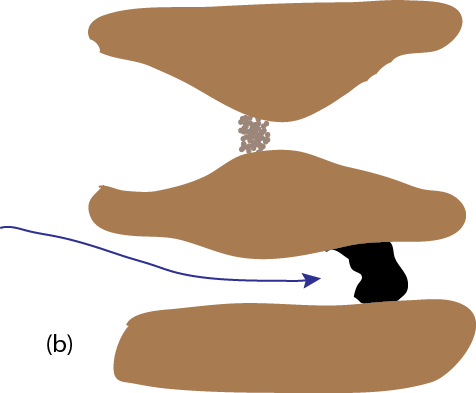}
\includegraphics[width=0.3\linewidth]{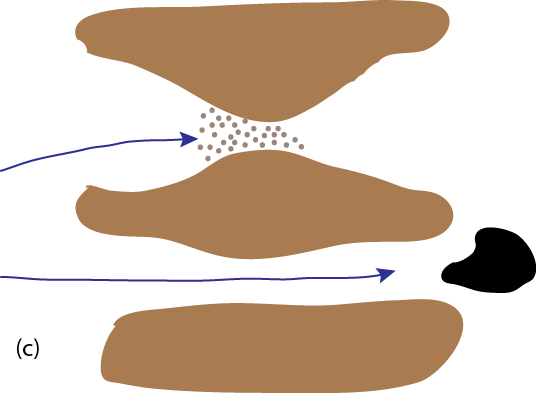}
\end{center}
\caption{(a) Water flowing into the upper channel, leading to the accumulation of particles and clogging of the pore. (b) Water flow is completely diverted towards to the lower channel, mobilizing the trapped oil. (c) The pressure release behind the accumulated particles leads to the restoration of the water filtration in the upper channel, eventually leading to unclogging of the pore.}
\label{fig.unc}
\end{figure*}

The unclogging of pore throats is a complex phenomenon to model.
For the transport of fine particles, as the ones released during low salinity waterflood, for example, an increase in the pressure above a certain critical value is enough to force particles through the constriction (or break them) \cite{civan2016}.
However, such effect requires that filtration of water between the accumulated particles takes place.
Since after a pore throat is clogged the water flow is diverted to a neighbouring channel, no significant filtration of water through the particles is expected.
We consider that unclogging occurs as a result of the pressure release from the water flow behind the accumulated particles when the water flow is completely diverted.
This is analogous to considering a relaxation time for the accumulation of particles.
After the pressure behind the clog is relieved due to complete diversion of the water flow, the particles start to move by diffusion in the static water.
Eventually, filtration of water is restored and the particles are advected through the unclogged pore.
A schematic is shown in Fig. \ref{fig.unc}.
According to this proposed mechanism, unclogging is mostly dependent on the volumetric concentration of accumulated particles.
For simplicity, we consider this dependence to be linear.

Note that our conceptual model considers that there are no interaction forces between the particles and the rock wall.
This is justified by the fact that polymer particles are negatively charged \cite{skauge2008}, such that attractive interactions between polymer particles and negatively charged sandstones, such as Berea \cite{nasralla2014}, are unlikely to occur.
Therefore, we consider that clogging of the pore is purely a mechanical effect.

For simplicity, at this stage we choose simple linear relations for the accumulation and removal functions, $\varphi(c_l,\sigma) = \varphi(c_l)$ and $\psi(c_l,\sigma)=\psi(\sigma)$, respectively and $\beta(c_l,\sigma)=\delta$ as the constant in Eq. (\ref{eq.r}).
Therefore, we consider the following rate equation
\begin{equation}
R 
= 
\delta (\phi k_l u_w c_l - k_r \rho_l\sigma)
\label{eq.rl}
\end{equation}
where $k_l$ and $k_r$ are constant rates accounting for clogging and unclogging and $u_w = |\vec{u}_w|$ is the magnitude of the water velocity.
One can see from Eq. (\ref{eq.rl}) that in a complete homogeneous medium, i.e., when $\delta=0$, log-jamming does not occur.
Equation (\ref{eq.rl}) is a modified version of the rate equation presented by Gruesbeck and Collins \cite{gruesbeck1982}, where they studied particle accumulation in a single-phase flow in a porous medium and neglected the detachment rate.
Moreover, the rock heterogeneity was not taken into account in their work.
Therefore, in comparison to the rate equation presented in \cite{gruesbeck1982}, we consider the effects of porosity $\phi$, water velocity $u_w$, heterogeneity $\delta$ and unclogging rate $-k_r\rho_l\sigma$.

It is worth to mention that not only the pore-size distribution is relevant for log-jamming, but also how the pores are connected.
For example, it is possible that many of the narrow pore throats are not connected to the larger pores, such that log-jamming is not favoured.
In fact, the connectivity of the pores is a key factor for determining the occurrence of log-jamming.
The hydrodynamic properties of the particles in the pore channel are also relevant with regards to determining whether unclog will occur or not.
Without a detailed knowledge of the water pathways inside the core, one can not expect to construct a complete model at this moment.

In the present model we neglect deposition and removal of particles at pore 
walls.
Log-jamming completely blocks a water pathway, whereas particle deposition lowers the area available for the water flux in the pore proportionally to the particle surface area, but it does not divert the water flow.
Therefore, the permeability reduction caused by log-jamming is expected to be more severe than the one caused by deposition.
We do account for a local lowering of porosity by particle accumulation, 
by the model
\begin{equation}
\phi = \phi_0 - \sigma,
\label{eq.por}
\end{equation}
where $\phi_0$ is the initial porosity.

%

Sequential clogging of pore channels will lead to the microscopic diversion of the water flow into neighbouring pore channels.
This may lead to the displacement of initially trapped-oil, i.e., an increase in the amount of oil available for mobilization.
In the next section we discuss this effect.

\subsection{Residual oil saturation}

After waterflood, the remaining oil is divided between capillary-trapped oil and oil which is mechanically trapped in some channels (as schematically shown in Fig. \ref{fig1}).
For capillary-trapped oils, the forces acting at the water-oil interface are weaker than the attachment force of oil at the pore wall, preventing its mobilization.
Mechanically-trapped oil, on the other hand, is oil located in pore channels where water does not flow, thus preventing its mobilization.
Capillary-trapped oil may be mobilized through an increase in the capillary number (a ratio between viscous force and interfacial tension between water and oil) \cite{shengBook}, while mechanically-trapped oil may be mobilized through a microscopic diversion of the water flow from the narrow channels into the larger pore channels containing trapped oil.

These two processes are fundamentally distinct, such that different strategies need to be used for each apart.
Techniques such as injection of surfactants \cite{shengBook} and microbial enhanced oil recovery \cite{landa2017}, aim at increasing the capillary number and recover capillary-trapped oil.
Injection of polymer particles, on the other hand, aims at recovering mechanically-trapped oil through clogging of the narrow water paths, leading to a microscopic diversion of the water flow into larger channels where oil is initially trapped.
One must note that microscopic diversion of the flow occurs at the pore-scale, which makes it distinct to diversion of flow to low-permeability areas (a core-scale process).

While these two processes are different from on the pore scale, on the core 
scale they are both manifested as an increase in the amount of oil available 
for mobilization.
This can be reproduced through a lowering on the residual oil saturation on the model.
In the literature there is an abundance of models for the residual oil saturation based on the capillary number, but none regarding the microscopic diversion of water flow, as far as the authors are aware.
For simplicity, we adapt a model taken from Li {\it et al} \cite{li2007}, hence considering the residual saturation as a function of the accumulated particles as
\begin{equation}
s_{or} 
=
\mbox{min}
\left(
s_{or}
,
s_{or}^{min} + (s_{or}^{ini} - s_{or}^{min})[1 + (A_1\sigma)^{A_2}]^{1/A_2-1}
\right),
\label{eq.sor}
\end{equation}
where $A_1$ and $A_2$ are constants which determine the rate at which residual oil saturation lowers,
$s_{or}^{min}$ is the absolute minimum value for the residual oil saturation and $s_{or}^{ini}$ is the initial residual oil saturation.
In the context of the present discussion, $s_{or}^{min}$ may be seen as the non-recovered capillary-trapped oil.

\subsection{Parametrizations}

The changes in porosity given by (\ref{eq.por}) induce changes in permeability.
There are many models in the literature relating permeability and porosity \cite{jacquin1964,adler1990}.
Ultimately, this relation depends on the detailed pore structure of the rock.
The accumulation of particles will lead to a linear decay in the porosity according to (\ref{eq.por}).
We utilize the following relation for the permeability \cite{hussain2013}
\begin{equation}
\frac{K}{K_0}
=
\frac{1}{1 + \gamma(\phi_0-\phi)}
=
\frac{1}{1 + \gamma\sigma}
,
\label{eq.kex}
\end{equation}
where $\gamma$ is a constant.

The phase mobilities $\lambda_\alpha = k_{r\alpha} / \mu_\alpha$ depends on the saturation $s_\alpha$ through the relative permeabilities $k_{r\alpha}(s_\alpha)$.
The viscosities are constant, as we assume that the flow is isothermal.
We consider a power-law of the Corey-Brooks type \cite{brooks1964} for the relative permeabilities
\begin{equation}
k_{ro} = (1-s_w^*)^{n_o},
\ \ \
k_{rw} = (s_w^*)^{n_w},
\label{eqS.krA}
\end{equation}
where $n_o$ and $n_w$ are constants, and $s_w^*$
\begin{equation}
s_w^* = \frac{s_w - s_{wi}}{1-s_{or} - s_{wi}}
\label{eqS.sstar}
\end{equation}
is the effective water saturation, with $s_{or}$ and $s_{wi}$ the residual oil saturation and irreducible water saturation, respectively.
The irreducible water saturation $s_{wi}$ does not change during injection of polymer particles, while the residual oil saturation $s_{or}$ changes according to Eq. (\ref{eq.sor}).

For the capillary pressure, we consider
\begin{equation}
p_c = - p_e\mbox{ln}(s_w^*),
\label{eq.cap}
\end{equation}
where $p_e$ is an entry pressure.
Note that Eq. (\ref{eq.cap}) does not account for hysteresis or dynamical effects \cite{duijn2016}.
In essence, we consider an equilibrium model for the capillary pressure.

If we consider the polymer particles as spheres with an average diameter of $d_p$, the diffusion coefficient may be calculated using the Stokes-Einstein relation
\begin{equation}
D = \frac{\tau K_B T}{2\pi\mu_w d_p},
\label{eq.d}
\end{equation}
where $K_B$ is the Boltzmann constant, $T$ is the absolute temperature and $\tau$ is the constant that measures the tortuosity of the flow.
In Eq. (\ref{eq.d}) we neglect dispersion of the polymer particles.

The set of Eqs. (\ref{eq.s})--(\ref{eq.sigmal}) models the transport of polymer particles in the water phase in a oil-water flow, considering the log-jamming rate as given by (\ref{eq.rl}).
By providing proper initial and boundary conditions, the problem can be solved numerically.

\section{Numerical results}
\label{sec.num}

The model presented in the previous section was implemented in MRST (Matlab Reservoir Simulation Toolbox) \cite{mrstbook}, an open-source reservoir simulator, which contains a vast range of data structures and computational methods implemented.
For our problem, we consider the EOR (Enhanced Oil Recovery) module and implement our model based on the polymer model presented there \cite{bao2017}.

The spatial domain is discretized in a simple $3$D cartesian grid with one block in both $y$ and $z$ directions, such that, essentially, a $1$D model is considered.
The fluxes are calculated on the cell faces, whereas the variables are cell-centered quantities.
Therefore, the model equations are discretized using a first-order implicit scheme in time and a standard two-point spatial scheme with upstream weighting, assuring mass conservation at each grid cell.
The discrete equations are then cast into residual form and a Newton method is employed to obtain the numerical solutions.
The standard tolerance for the Newton solver embedded in MRST is of $10^{-8}$.

\subsection{Model capabilities}
Before we compare the results obtained from the present model with the experimental data given in \cite{spildo2009}, we present some illustrative numerical results in order to highlight the main properties of the model.
For such, we neglect gravity and capillary effects and consider a domain of length $L = 2~m$, area $4\times10^{-2}~m^2$, initial permeability $K_0 = 1~\mu m^2$ and initial porosity $\phi_0 = 0.3$.
The diffusion coefficient $D$ is calculated for coreflooding occurring at \cite{spildo2009} $T = 358.15~K$, water viscosity $\mu_w = 0.29~centi\cdot poise$, polymer particles of diameter $d_P=40~\mu m$ and tortuosity $\tau=5$.
We have the following parameters (the values of the reaction rates were chosen from dimensional analysis)
\begin{table}[h!]
	\centering
	\caption{Parameters used for numerical calcuations.}
	\label{tab1}	
	\begin{tabular}{|c |c|}
	\hline
	$D = 3\times10^{-11}~m^2/s$& $\mu_o  = 3.6~centi\cdot poise$\\
	$k_l = 10^3~m^{-1}$        & $\rho_l = 1100~kg/m^3$\\
	$k_r = 10^{-6}~s^{-1}$     & $\delta = 1$    \\
	$s_{or}^{ini} = 0.5$       & $n_w    = 2.5$  \\
	$s_{wi} = 0.05$            & $n_o    = 3.0$  \\
	\hline
	\end{tabular}
\end{table}

The fitting parameters are given by $A_1 = 10, A_2 = 1\times10^4$ and $\gamma = 2\times10^3$.
The backpressure is kept fixed at $p(L,t) = 10^5~Pa$, whereas the injection rate of water is kept at $1.67\times10^{-6}~m^3/s$ at $x=0$.
Initial and boundary conditions are given respectively by
\begin{align}
	t = 0 &:
	\ \
	s_w(x,0) = 0.06, 	\ \ \	c_l(x,0) = 0.
	\\[5pt]
	x = 0, t > 0 &:
	\ \
	s_w(0,t) = 1, 	\ \ \	c_l(0,t) = 1~(kg/m^3)~H(t-t_l)(1 - H(t-t_w)).
	\label{eq.IBC}
\end{align}
We start injecting particles at a mass concentration of $1~kg/m^3$ at $t = t_l$, the time at which oil production due to injection of pure water ceases.
When $t=t_w$, we stop injecting polymer particles and perform a postwater flush for $t>t_w$.
The time $t_w$ is when oil production due to polymer particle injection decrease.
Therefore, the water viscosity (in $centi.poise$) is given by \cite{spildo2009}
\begin{equation}
\mu_w
=
\left\{
\begin{array}{l}
	0.29, \ \ t<t_l \ \mbox{and} \ t>t_w,
	\\
	0.90, \ \ t_l<t<t_w,
\end{array}
\right.
\end{equation}

\begin{figure*}[h!]
\hspace{-0.5cm}
\includegraphics[width=0.5\linewidth]{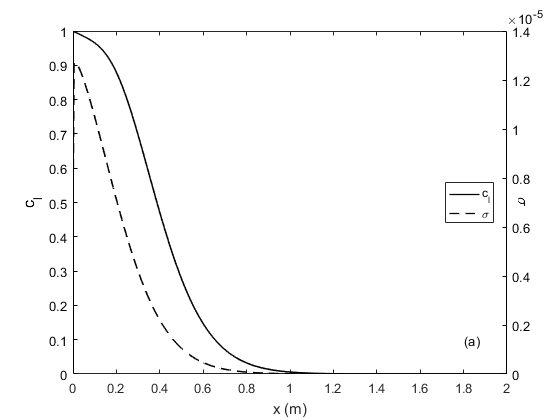}
\includegraphics[width=0.5\linewidth]{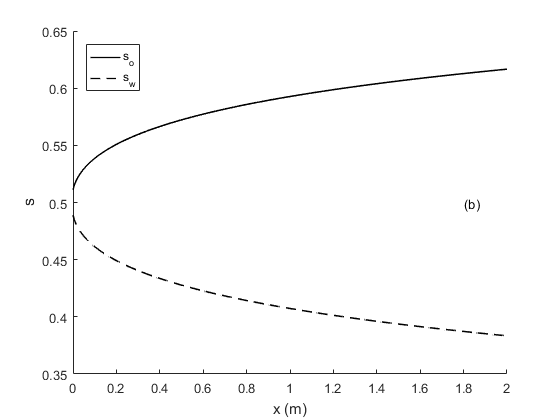}
\caption{(a) Mass concentrations of polymer and volumetric concentration of accumulated particles.
(b) Oil and water saturation. Flow direction is from left to right.}
\label{fig.lpsBT}
\end{figure*}

In Fig. \ref{fig.lpsBT}(a), we show the profiles of $c_l$ and $\sigma$ for the beginning of the particle front propagation.
The polymer particles start to propagate downstream (from left to right) as a smooth front, connecting the initial value $c_l = 0$ to some value $c_l < c_l^{inj}$, which is smaller than the injection value because of the particle accumulation in the upstream side of the front.
At the time step shown in Fig. \ref{fig.lpsBT}(a), the accumulation of particles is low, such that the flow is essentially non-disturbed.
This can be seen in Fig. \ref{fig.lpsBT}(b), where the profiles for oil and water saturation are shown for the same time step.

\begin{figure*}[h!]
\begin{center}
\hspace{-0.5cm}
\includegraphics[width=0.5\linewidth]{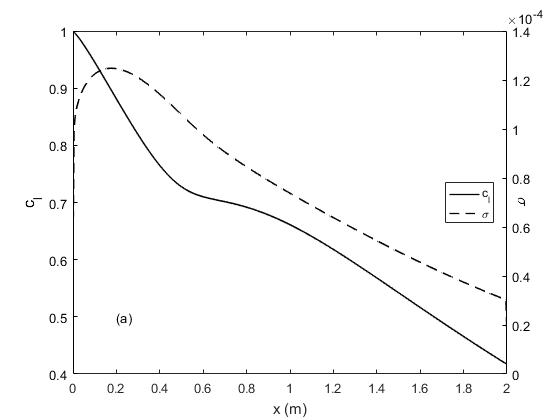}
\includegraphics[width=0.5\linewidth]{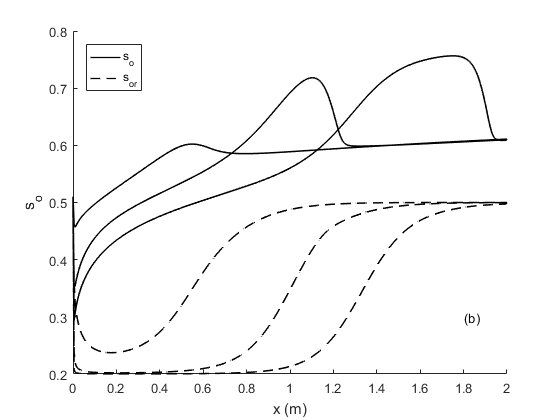}
\caption{
(a) Mass concentrations of polymer and volumetric concentration of accumulated particles when trapped oil begins to be mobilized.
(b) Oil saturation at different time steps, showing the formation of a dual-shock structure.}
\label{fig.so}
\end{center}
\end{figure*}

The accumulation of particles in the upstream side of the particle front eventually is large enough to clog some pore channels, leading to the mobilization of initially trapped oil through a lowering on the residual oil saturation according to Eq. (\ref{eq.sor}).
The newly mobile oil is displaced by the water flow upstream, which creates non-monotone oil saturation and particle concentration profiles.
The oil front eventually evolves into two shock fronts.
The slower shock, which is preceded by a rarefaction wave, is located at the point where the residual oil saturation change from $s_{or}^{min}$ to $s_{or}^{ini}$.
The faster shock is a typical Buckley-Leverett oil-water shock, but jumping from a high oil saturation to a low oil saturation downstream.
These features can be seen in Figs. \ref{fig.so}(a) and (b).
The dual shock structure shown in Fig. \ref{fig.so}(b) is similar to the one present in surfactant injection, and therefore their velocities can be calculated using standard Buckley-Leverett theory \cite{pope1980}.
Note that the transition from $s_{or}^{min}$ to $s_{or}^{ini}$ is smooth, such that the slower shock is diffuse.

\begin{figure*}[h!]
\begin{center}
\hspace{-0.5cm}
\includegraphics[width=0.5\linewidth]{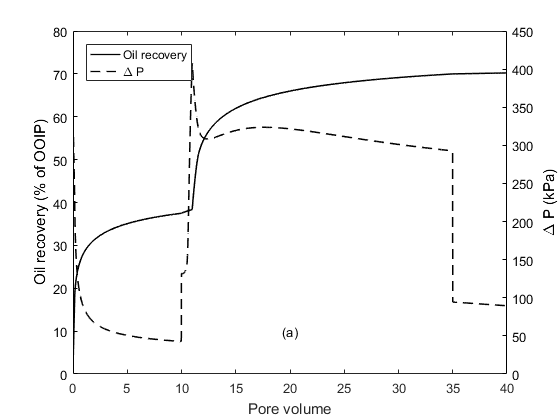}
\includegraphics[width=0.5\linewidth]{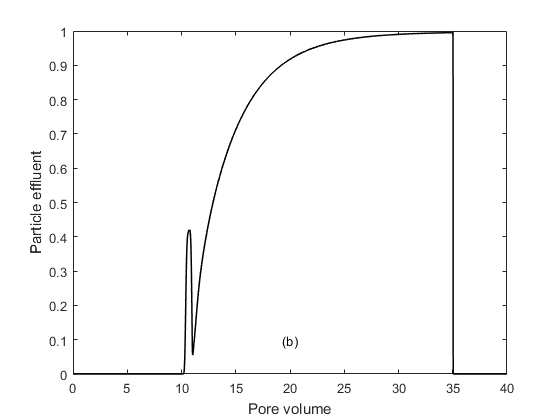}
\caption{
(a) Oil recovery and pressure drop.
(b) Particle effluent normalized with respect to the injected concentration. The sudden decrease in the particle effluent around $11$ pore volumes occurs when the oil front reaches the outlet.}
\label{fig.oilEx}
\end{center}
\end{figure*}

When the oil front reaches the outlet, recovery increases significantly.
In Fig. \ref{fig.oilEx}(a) we present the oil recovery (in percentage of original oil initially in place) and the pressure drop across the core.
We inject pure water for a period of $10$ pore volumes, after which we begin to inject water with polymer particles for a period of $25$ pore volumes, before postwashing it with pure water for $5$ pore volumes.
Accumulation of particles behind the particle front leads to displacement of initially trapped oil.
The final amount of oil recovered is determined by the strength of the reaction rate and by the minimum residual oil saturation $s_{or}^{min}$.
According to our assumptions, $s_{or}^{min}$ corresponds to the oil that remains capillary-trapped after mobilization of trapped oil through log-jamming and it can not be recovered through particle accumulation.
The lowering in the residual oil saturation in this case was of $52.36\%$.

The pressure drop along the core increases as production of oil takes place, decreasing to a steady value when only the injected phase is produced at the outlet.
There is a new decrease in the pressure drop when we cease injection of polymer particles and replace it by injection of pure water (postwater flush) due to viscosity decrease.
As injected particles begin to clog pore channels, permeability reduction takes place and the pressure drop increases in comparison with pure water flood.
The pressure drop along the core is inversely proportional to the permeability $K$ and the total mobility $\lambda = \lambda_w+\lambda_o$.
If we neglect capillary pressure and gravity, the ratio between the pressure drops after postwater flush $\Delta P_{pw}$ and waterflood $\Delta P_{wf}$ is given by
\begin{equation}
\frac{\Delta P_{pw}}{\Delta P_{wf}}
=
\left(\frac{K^0}{K^F}\right)
\left(\frac{\lambda^0}{\lambda^F}\right),
\label{eq.dp}
\end{equation}
where the superscripts $0$ and $F$ denote initial and final values, respectively.
Accumulation of particles leads to a decline in the core permeability, while lowering in the residual oil saturation leads to an increase in the mobility.
Thus, these two processes have opposite effects in the overall pressure drop.
In the present case, we have a pressure drop ratio of $2.00$, a permeability reduction of $62\%$ and a mobility ratio of $1.32$.
During the postwater flush, only unclogging occurs, as particles are no longer injected into the core.
This causes the slight decrease in the pressure drop between $35$ and $40$ pore volumes, as seen in Fig. \ref{fig.oilEx}(a).

In Fig. \ref{fig.oilEx}(b) we show the particle effluent as a ratio of the concentration of particles at the outlet and the injected particles.
The particle front travels ahead of the oil front, carried by the water flow.
Under the conditions given in Table \ref{tab1}, the particle front reaches the outlet prior to the oil front (first increase shown in Fig. \ref{fig.oilEx}(b), right after $10$ pore volumes).
When oil starts to be produced at the outlet (at around $11$ pore volumes), the effluent of particles decrease abruptly because of the lower water content, increasing again as oil production ceases.
After production of oil ceases, the effluent of particles is mainly determined by the competition between clogging and unclogging.
An equilibrium state is achieved when $R=0$, i.e., when clogging and unclogging balance each other according to Eq. (\ref{eq.rl}).
From Fig. \ref{fig.oilEx}(a), one can see that for times after around $17$ pore volume injected, the pressure drop across the core begins to lower.
This is an effect of unclogging becoming more relevant than clogging, i.e.,  $R<0$.
In this case, one can see from Fig. \ref{fig.oilEx}(b) that between $25$ and $35$ of pore volumes injected, the particle effluent is complete.

\subsection{Comparison with experimental results}

In this Section we compare numerical results with the experimental data taken from \cite{spildo2009}.
The experiments described in \cite{spildo2009} were conducted in five different cores, from which we take three for comparison\footnote{For the other two cores: one was not stabilized during polymer particle injection and the other achieved pressure drops one order of magnitude higher than the other cores, such that we considered those two cores to be outside the parametric range for comparison considered here.}, with different rock properties.
After the initial preparation of the cores, the experiments had three stages: waterflood at a rate of $1.67\times 10^{-9}~m/s$, waterflood at a rate of $1.67\times 10^{-8}~m/s$, injection of polymer particles with water at a rate of $1.67\times 10^{-8}~m/s$ and a post water flush at the same rate (water with no particles).
We fix some parameters and vary the available rock properties in order to compare our results with the reported experimental results.
The varying parameters will be the initial rock permeability $K_0$, initial porosity $\phi_0$, initial residual oil saturation $s_{or}^{ini}$, irreducible water saturation $s_{wi}$, rock heterogeneity $\delta$ and unclogging constant $k_r$.
The rock heterogeneity is calculated as explained in Section \ref{sec.log} using the available data for the pore-size distribution of each representative core (see Appendix A).
The unclogging constant is varied in order to adjust the final pressure drop ratio.

The cores have similar sizes and for simplicity we consider the same length of $L=0.2~m$ and area $9\times10^{-4}~m^2$ for the three cores.
The specific values for each core are given in Table \ref{tabCores}.
The entry pressure $p_e$, necessary for the capillary pressure, is given by $p_e = 10^3~Pa$, whereas the clogging constant is given by $k_l=2\times10^{4}~m^{-1}$.
It is worth to mention that the backpressure $p(L,t)$ and the entry pressure $p_e$ are one order of magnitude smaller than the real values considered in the experiments \cite{spildo2009}.
This is due to numerical restrictions: we are considering a small domain, such that high values for the pressure would cause variations too large to be handled by the numerical solver.
Otherwise specifically stated, the remaining parameters are the same as the ones given in Table \ref{tab1}.

\begin{table}[h!]
	\centering
	\caption{Data varied from core to core.}
	\label{tabCores}
	\begin{tabular}{|c |c |c |c|}
          \hline
                    & Core A & Core B & Core C \\
          \hline
	$K_0 (\mu m^2)$ & $0.90$   & $0.50$   & $0.30$  \\
	$\phi_0$        & $0.33$   & $0.32$   & $0.33$  \\
	$s_{or}^{ini}$  & $0.51$   & $0.30$   & $0.32$  \\
	$s_{or}^{min}$  & $0.20$   & $0.17$   & $0.19$  \\
	$s_{wi}$        & $0.06$   & $0.18$   & $0.24$  \\
		 \hline
	$\delta$        & $0.2315$ & $0.0540$ & $0.0354$\\	
	$k_r (s^{-1})$  & $2.0\times10^{-8}$    & $1.2\times10^{-4}$ & $4.0\times10^{-4}$ \\
          \hline
	\end{tabular}
\end{table}

Initial and boundary conditions are given respectively by
\begin{align}
	t = 0 &:
	\ \
	s_w(x,0) = s_{wi}+0.01, 	\ \ \	c_l(x,0) = 0.
	\\[5pt]
	x = 0, t > 0 &:
	\ \
	s_w(0,t) = 1.0, 	        \ \ \	c_l(0,t) = 1~(kg/m^3)~H(t - t_l)(1-H(t - t_w)).
	\label{eq.IBCB}
\end{align}
The initial condition is such that the core is mostly oil-saturated prior to waterflood.
Injection of polymer particles along with water only occurs after some time $t_l$, which is chosen as the time when oil production ceases after the second waterflood.
When oil production ceases during the stage of particles injection (which occurs at a time $t_w$), we start injecting pure water (post water flush).
As mentioned previously, we consider two stages of waterflood before the injection of polymer particles.
The rates considered as boundary conditions are the same as the ones utilized in the experiments conducted in \cite{spildo2009}.

\begin{figure*}[h!]
\begin{center}
\hspace{-0.5cm}
\includegraphics[width=0.5\linewidth]{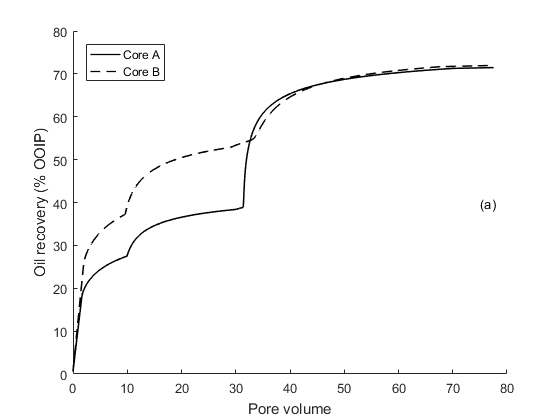}
\includegraphics[width=0.5\linewidth]{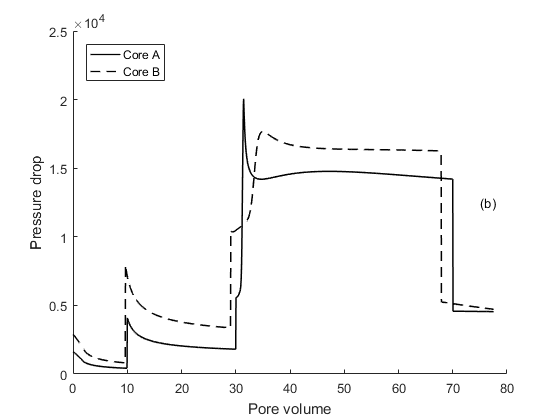}
\includegraphics[width=0.5\linewidth]{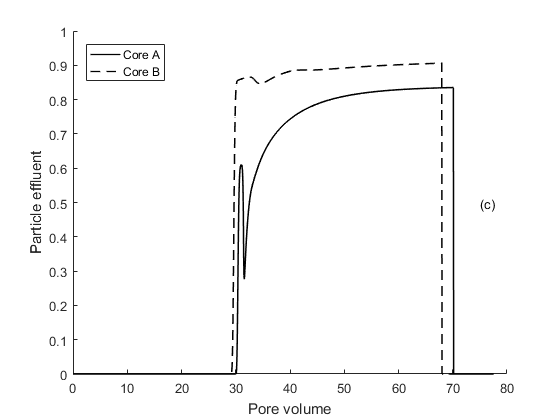}
\caption{(a) Oil recovery curves for cores A and B. (b) Pressure drop across the cores A and B. The peaks correspond to oil production and the plateaus to water breakthrough. (c) Particle effluent for cores A and B.}
\label{fig.oilAB}
\end{center}
\end{figure*}

Results for oil production curves for cores A and B are show in Fig. \ref{fig.oilAB}(a).
We inject water at a rate of $1.67\times10^{-9}~m/s$ for $10$ pore volumes, then increase the injection rate up to $1.67\times10^{-8}~m/s$ and keep it for $20$ pore volumes, when production of oil ceases.
Then we start injecting water with polymer particles with a mass concentration of $1~kg/m^3$ according to (\ref{eq.IBCB}).
Injection of polymer particles is sustained for $40$ pore volumes, when oil production ceases.
The final stage is a postwater flush for $10$ pore volumes.
These injection periods are longer than the ones reported in \cite{spildo2009} because we intended to guarantee that oil production ceased.
Since the injection rate is the same for both cores, but core B has a lower permeability, the recovery of oil during waterflood is larger for core B.
However, the increase in oil recovery due to polymer particle injection is smaller in core B than in core A.
This is a result from the fact that accumulation of particles depends on the heterogeneity of the core as explained previously.
Cores with a lower degree of heterogeneity tend to have a slower accumulation of particles, leading to a smaller oil mobilization and therefore resulting in a smaller increase in oil recovery.

During production, the pressure drop along the core increases sharply, followed by a decline after production ceases, when it establishes at a constant value.
In Fig. \ref{fig.oilAB}(b) we present the pressure drop along the core for cores A and B.
The maximum pressure drop occurs at the moment of oil production due to polymer particles injection.
The exact value of the maximum pressure drop depends not only on the amount of particles at the effluent, but also on the rock permeability.
Since reaction is stronger in core A, more particles are accumulated inside it.
This can be seen from Fig. \ref{fig.oilAB}(c), where the particle effluent is shown for both cores.

If we denote $\Delta P_{pw}$ as the overall pressure drop along the core after the postwater flush and $\Delta P_{wf_2}$ as the overall pressure drop along the core after the $2$nd waterflood, we can calculate the ratio $IPR = \Delta P_{pw}/\Delta P_{wf_2}$.
This quantity accounts for the overall effects of permeability reduction and mobility increase due to clogging and mobilization of initially-trapped oil.
In Table \ref{tab3} we compare experimental and simulation results for the $IPR$ and the lowering in the residual oil saturation.
We also present the calculated permeability reduction for each core.
As one can see, there is a trend for a lower increase in oil recovery and lower $IPR$ for the more globally homogeneous core (lower value of $\delta$).
The more heterogeneous core tend to have more accumulation of particles, which will lead to a higher degree of microscopic diversion of the flow, thus mobilizing more initially-trapped oil.
This experimentally-observed trend can be captured by considering a reaction rate depending on the heterogeneity, as was done in the present work.

\begin{table}[h!]
	\centering
	\caption{Comparison between experimental and numerical results.}
	\label{tab3}
	\begin{tabular}{|c |c |c |c|}
          \hline
                                       & Core A& Core B& Core C \\
          \hline
		  $IPR$ \cite{spildo2009}      & $2.60$ & $1.40$ & $1.00$ \\
		  $IPR$ (numerical)            & $2.51$ & $1.40$ & $1.06$ \\
          \hline
   		  $\%$ reduction in $s_{or}$ \cite{spildo2009}  & $61.0$  & $42.0$ & $41.0$ \\
   		  $\%$ reduction in $s_{or}$ (numerical)        & $53.8$  & $40.5$ & $37.3$ \\
          \hline
   		  $\%$ permeability reduction                   & $66.3$  & $41.6$ & $20.4$ \\		  
          \hline
	\end{tabular}
\end{table}

Note that the reduction in residual oil saturation for cores B and C are similar, although their $IPR$ are significantly different.
An experimental value of $IPR = 1$ for core C indicates that the final permeability reduction is very low.
This suggests that after oil mobilization takes place, almost total unclogging of the pores occurred.
In our model we adjusted the value of the unclogging constant $k_l$ for each core in order to match experimental data.
By performing a proper analysis of the problem at the pore scale, one should be able to predict the qualitative dependence of the reaction rate constants $k_r$ and $k_l$.
Therefore, $k_r$ and $k_l$ will be parameters that depend on the mechanical-physical interactions between the oil-water flow, particle transport and rock geometry at the pore scale.
%

\section{Conclusions}

We developed a simple model for the transport of polymer particles in an oil-water flow in a porous medium, which takes into account the clogging/unclogging of pore throats.
We show that recent experimental results \cite{spildo2009,spildo2010} can be explained by considering a dynamic alteration of the permeability and a macroscopic degree of the microscopic core heterogeneity.
Nevertheless, a detailed derivation of the reaction rates at the pore-scale must be conducted in order to better clarify the complicated coupled physics in this problem.

The transport of polymer particles in a oil-water flow in porous media is a complex problem.
In particular, we point to the difficulty in capturing all physical meaningful effects with a heuristic model, without resorting to large-scale experimental fitting.
Moreover, in the limit of severe permeability decrease, i.e., if $K\rightarrow0$, the problem will be degenerated, such that the Newton solver embedded in MRST might not have an efficient convergence.
In this case, different linearization schemes must be implemented in order to efficiently numerically solve the problem \cite{radu2015,list2016}.
In order to compare with experimental results, we considered essentially a one-dimensional domain.
For reservoirs, the large heterogeneities existent may favour formation of fingers, which compromise oil recovery.
Therefore, a stability analysis would be relevant for reservoir applications \cite{musuuza2009,musuuza2011}.
A future work will also consider a more detailed derivation of the clogging/unclogging model.

\appendix

\section{Heterogeneity factor}

\begin{figure*}[h!]
\begin{center}
\hspace{-0.5cm}
\includegraphics[width=0.5\linewidth]{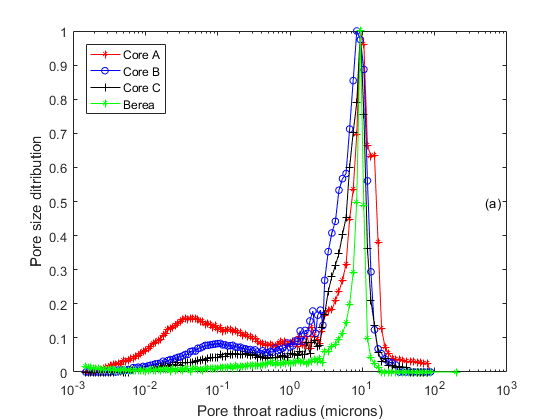}
\includegraphics[width=0.5\linewidth]{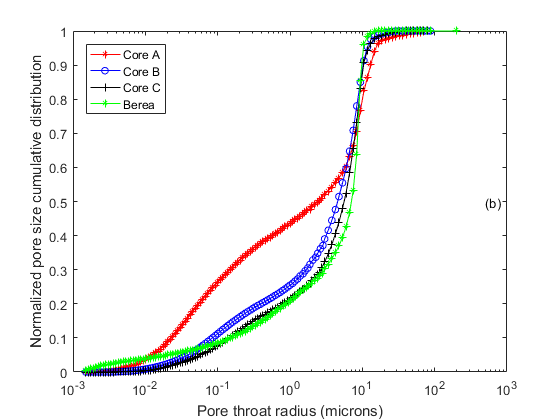}
\caption{(a) Normalized pore size distribution and (b) normalized cumulative distribution function.}
\label{figB.psd}
\end{center}
\end{figure*}

As mentioned previously, a key aspect for the accumulation of particles is the heterogeneity of the core.
In Fig. \ref{figB.psd}(a) we show the normalized pore size distributions and in (b) their respective normalized cumulative distribution functions, for cores A, B and C.
We also plot the distribution functions for the Berea core, which is the homogeneous core utilized for comparison.
Experiments with polymer particle injection have shown negligible increase in oil recovery for Berea cores \cite{skauge2010}.
Since Berea is a fairly homogeneous core, this supports our claim that heterogeneity plays a significant role in microscopic diversion.
Therefore, we chose Berea as a representative homogeneous core for the calculation of the heterogeneity factor $\delta$.

\begin{figure*}[h!]
\begin{center}
\hspace{-0.5cm}
\includegraphics[width=0.5\linewidth]{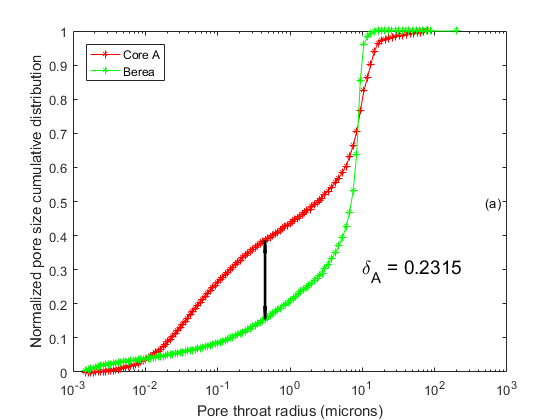}
\includegraphics[width=0.5\linewidth]{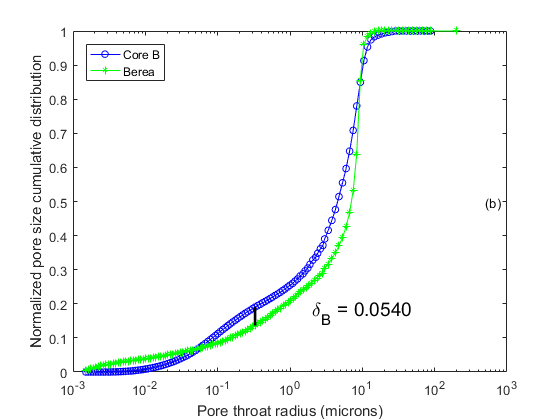}
\includegraphics[width=0.5\linewidth]{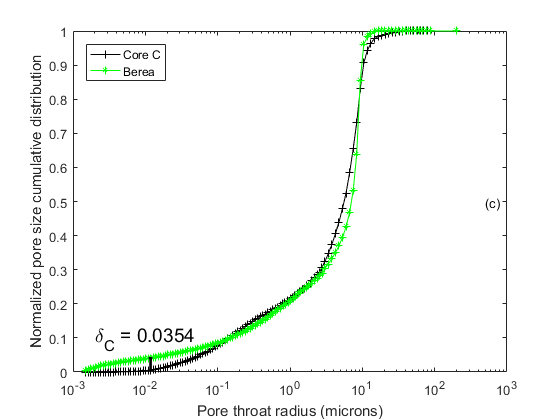}
\caption{Heterogeneity factor for cores (a) A, (b) B and (c) C.}
\label{figB.diff}
\end{center}
\end{figure*}

In Figs. \ref{figB.diff}(a) to (c) we present the comparison between cores A to C and the Berea, with their respective values for $\delta$.
In essence, $\delta$ is a measurement of how far from a homogeneous core (Berea) the given sample is.
Larger values of $\delta$ gives a high heterogeneity, indicating that the core is more suitable for oil recovery enhancement through polymer particle injection.
For convention, pores with radius $r < 1~micron$ are considered in the microscopic region, whereas pores with radius $> 1~micron$ are at the macroscopic region.
We consider the maximum separation between distributions $\delta$ in the microscopic region.
In other words, we neglect the variations in pore sizes in the macroscopic region ($r>1~micron$).
This is justified by the fact that particle-carrying water flow only clogs a pore when the large-to-narrow pore throat ratio is large.
In other words, a flow change from two pores with different throat radius does not permit clogging if both pores are in the macroscopic region.

\end{document}